\documentclass[aps,prl,twocolumn,showpacs,amsmath]{revtex4}
\usepackage{graphicx}

\begin{document}

\title{Titanium-decorated carbon nanotubes: 
a potential high-capacity hydrogen storage medium }

\author{T. Yildirim$^1$ and S. Ciraci$^{2}$}

\affiliation{$^{1}$NIST Center for Neutron Research, National
Institute of Standards and Technology, Gaithersburg, MD 20899\\
$^{2}$Physics Department, Bilkent University, 
06800 Bilkent, Ankara, Turkey}

\begin{abstract}
We report a first-principles study, 
which demonstrates that a single Ti atom coated on a 
single-walled nanotube (SWNT)  binds up 
to four hydrogen molecules. The first H$_2$ adsorption 
is dissociative with no energy barrier while other three 
adsorptions are molecular with significantly elongated H-H bonds. 
At high Ti coverage we show that a  SWNT can strongly 
adsorb up to 8-wt\%  hydrogen. 
These results  advance our fundamental understanding 
of dissociative adsorption of hydrogen in nanostructures 
and suggest new routes to better storage and catalyst materials. 
\end{abstract}

\pacs{61.46.+w,68.43.-h,84.60.Ve,81.07.De} 
\date{Submitted to PRL Nov. 2004, Accepted March 2005, to appear in May 2005}
\maketitle

Developing safe, cost-effective, and practical means of storing hydrogen 
is crucial for the advancement of hydrogen and fuel-cell 
technologies\cite{science-review}. 
The current state-of-the-art is at an impasse in providing any material
 that meets a storage capacity of 6-wt\% or more required for practical 
 applications\cite{science-review,chan,tada,miura,gang,dubot,lee,gulseren}. 
 Here we report 
 a first-principles computation of the interaction between 
 hydrogen molecules and transition metal atoms adsorbed 
 on carbon nanotubes. 
 Our results are quite remarkable and unanticipated. We found 
 that a single Ti-atom adsorbed  on a SWNT can strongly bind up to 
 four hydrogen molecules. Such an unusual and complex bonding
 is generated by the concerted interaction among H, Ti, and
 SWNT. 
 Remarkably, this adsorption occurs with no energy barrier. 
 At large Ti coverage we show that a (8,0) SWNT can store hydrogen 
 molecules up to 8-wt\%, exceeding the minimum requirement of 
 6-wt\% for practical applications. Finally, we present high 
 temperature quantum molecular dynamics simulations showing that 
 these systems are stable and indeed exhibit associative 
 desorption of H$_2$ upon heating, another requirement for 
 reversible storage. 
 
 Recent experiments\cite{zhang-apl,zhang-cpl} and 
 calculations\cite{gulseren-prl,yang,durgun,dag}
 suggest that it is possible to coat 
 carbon nanotubes uniformly with Ti atoms without metal
segregation problems\cite{segregation}. 
Here we show that such Ti-coated
carbon nanotubes exhibit remarkable hydrogen absorption 
properties. Below we will present our results in detail
for a (8,0) nanotube and briefly for 
four  armchair (n,n) (n=4,5,6, and 7) and five zigzag 
(n,0) (n=7,8,9,10,11, and 12) nanotubes.

A single Ti atom on 
 an (8,0) SWNT has a magnetic ground state with S=1 and a binding 
 energy of 2.2 eV; this will serve as our reference system, 
 denoted {\bf t80Ti}. 
 In order to determine different reaction paths and products between 
 H$_{2}$ and {\bf t80Ti}, we have carried out a series of single-energy 
 calculations as H$_2$ molecules approaches {\bf t80Ti} and when 
 there are large enough forces acting on H$_2$ molecules, 
 we let the atoms evolve according to the quantum mechanical 
 forces obtained from density functional theory (DFT) 
calculations\cite{castep}. 
 We used the conjugated-gradient (CG) minimization and optimized both 
 the atomic positions and the c-axis of the tube. 

The energy calculations were performed within the plane-wave 
implementation\cite{castep} of the generalized gradient approximation (GGA-PBE)\cite{pbe} 
to DFT. We used Vanderbilt ultra-soft pseudopotentials\cite{usp} treating the 
following electronic states as valence: Ti: $3s^{2}3p^{6}3d^{2}4s^{2}$, 
C: $2s^{2}2p^{2}$ and H: $1s$. 
The Monkhorst-Pack special k-point scheme\cite{kpts} is used with 
0.025 $\AA^{-1}$ k-point spacing resulting in 5 k-points along the tube axis. 
The cutoff energy of 350 eV is found to be enough for total 
energies to converge within 0.5 meV/atom.  The calculations are 
carried out in a tetragonal supercell geometry of $20 \AA \times 20 \AA \times c$ 
where $c$ is taken to be twice the lattice constant of SWNT along its axis. 

%
\begin{figure}
\includegraphics[scale=0.70,angle=0]{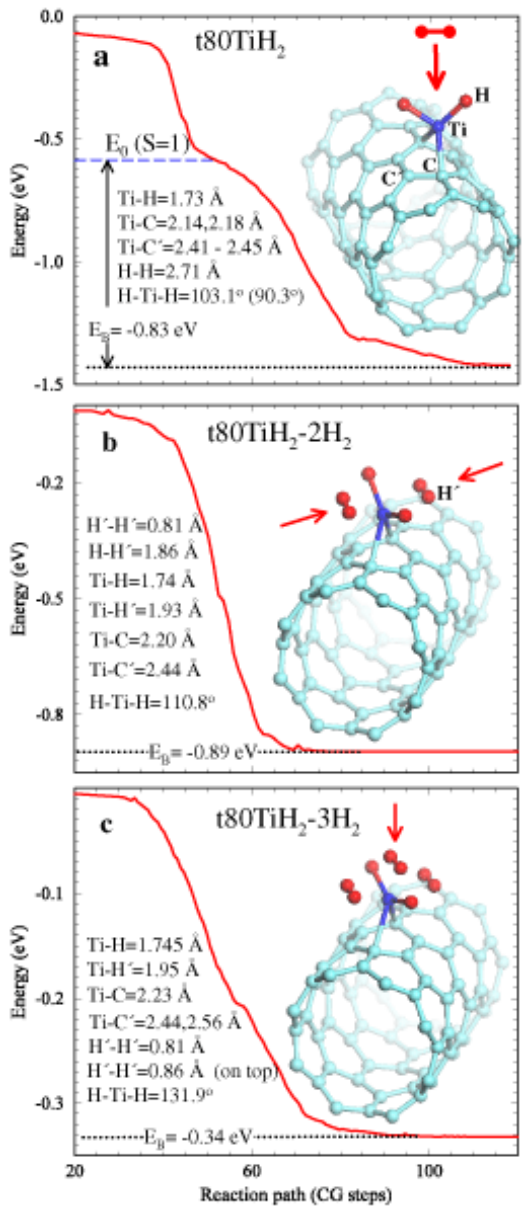}
\caption{
Energy versus reaction paths for successive 
dissociative and molecular adsorption of 
H$_{2}$ over a single Ti-coated (8,0) nanotube. 
(a)$ H_{2} + t80Ti \rightarrow t80TiH_{2} $. 
(b) $2H_{2}+ t80Ti H_{2} \rightarrow t80TiH_{2}-2H_{2}$. 
(c) $H_{2}+ t80TiH_{2}-2H_{2} \rightarrow t80TiH_{2}-3H_{2}$. 
The zero of energy is taken as the sum of the energies of two reactants. 
The relevant bond distances and   binding energies ($E_{B}$)
are also given.
}
\label{fig1}
\end{figure}

Figure 1a shows the energy variation from {\it non-spin-polarized}
calculations as a single H$_2$ molecule approaches 
{\bf t80Ti}. 
The energy first decreases slowly as the hydrogen gets closer to the nanotube 
and Ti. However, as the charge overlap gets large, the H$_{2}$
 molecule is attracted 
towards the Ti atom with a sudden decrease in the energy. At this point, 
the H$_{2}$  molecule is still intact with a significantly increased H-H bond 
length of 0.86 \AA. The second sudden decrease in energy is achieved by 
dissociating the H$_{2}$ molecule into two H atoms. At this point, the H-H 
distance increases from 0.86 \AA \; to 2.71 \AA. 
The interaction between H$_{2}$ and 
{\bf t80Ti} is always attractive and therefore H$_{2}$ is 
absorbed onto a Ti atom 
without any energy barrier. The final geometry is shown in the inset to 
Fig.~1a, with relevant structural parameters given\cite{animations}. 
In order to calculate the binding energy for this dissociative 
adsorption, we calculate the total energies  of the initial 
t80Ti and H$_2$ state (dashed line in Fig.~1a) and the final 
t80TiH$_2$ state (dotted line in Fig.~1a) from {\it spin-polarized}
calculations. We obtain the binding energy to be 0.83 eV (Fig.~1).
We note that {\it spin-polarized} calculation lowers the total
energy by about 0.6 eV with respect to non-spin-polarized calculations
and yields a triplet magnetic ground state (i.e. S=1) for the initial
t80Ti and H$_2$ system.
However once the hydrogen molecule is attached to Ti, the system
is non-magnetic and spin-polarized calculations are not necessary.


Remarkably, it is also energetically favorable for the TiH$_{2}$ group 
to complex with additional hydrogen molecules.  Fig.~1b shows the energy 
variation as two hydrogen molecules approach the Ti atom; one from each 
side of the TiH$_{2}$  group. 
As in the case of single adsorption, the energy 
always decreases, first slowly and later very rapidly at which point 
both hydrogen molecules are strongly attached to the {\bf t80TiH$_{2}$} system. 
We denote the final product as {\bf t80TiH$_{2}$-2H$_{2}$}, 
which is shown in Fig. 1b. 
The total energy change upon adsorption is 
about 0.89 eV (i.e. 0.45 eV/H$_{2}$). 
Unlike the first adsorption, the two hydrogen molecules are in intact but 
with a rather elongated bond length of 0.81 \AA. This 10\% increase is 
rather reminiscent of the elongated H-H bonds observed in the 
metal-dihydrogen complexes first synthesized by Kubas\cite{kubas,kubasbook}.

Fig.~1c shows the energy evolution when a fourth hydrogen molecule 
approaches the {\bf t80TiH$_{2}$-2H$_{2}$}  system from the top. 
The energy again 
decreases continuously, indicating a zero-energy barrier. 
The final product, denoted as {\bf t80TiH$_{2}$-3H$_{2}$}, 
is shown in the inset 
with the relevant structural parameters. The energy gained by 
the fourth adsorption, 0.34 eV/H$_{2}$, is slightly smaller than for 
the other cases but is still substantial. The H-H distance of 
the top H$_{2}$ is 0.86 \AA.  Several attempts to add a fifth hydrogen 
molecule at a variety of positions failed, suggesting 
a limit of 4 H$_{2}$/Ti.

The final optimized structures shown in Fig.~1 
need not be the global minimum. 
Among many different isomers tried for the 
four-H$_{2}$ system, we found a very 
symmetric configuration (denoted as {\bf t80Ti-4H$_{2}$}) 
that is 0.1 eV lower in 
energy than {\bf t80TiH$_{2}$-3H$_{2}$}  (Fig. 2a).
 Here all four hydrogen molecules stay 
intact and benefit equally from bonding with the Ti atom.
 The average 
H-H bond distance is about 0.84 \AA $\;$ and each molecule 
has an excess charge of about 0.15 e. 
The average binding energy per H$_2$ molecule is 0.54 eV, 
which suggests that the 
bonding is an unusual combination of chemisorption and physisorption.

To test system stability, we performed high-temperature quantum molecular 
dynamics (MD)
simulations on {\bf t80TiH$_{2}$-3H$_{2}$} \cite{animations}. 
In total, we have carried out 
 a 1.5 ps MD simulations with a Langevin thermostat for temperatures 
 ranging from 200 K to 900 K. We observe an associative desorption of 
 a H$_2$ molecule (i.e, two H atoms come together to form H$_{2}$ and 
 leave the system) 
 at around 800 K\cite{animations}. 
 While a 1.5 ps time MD simulation is 
 already computationally quite costly, it is not long enough to 
 get statistically meaningful values for the temperature. 
 However, it does suggest that the system is quite stable and that 
 it is possible to extract the H$_{2}$ molecules without 
 breaking the Ti-C 
 bonding or removing the TiH$_{2}$ from the nanotube. 

\begin{figure}
\includegraphics[scale=0.50,angle=0]{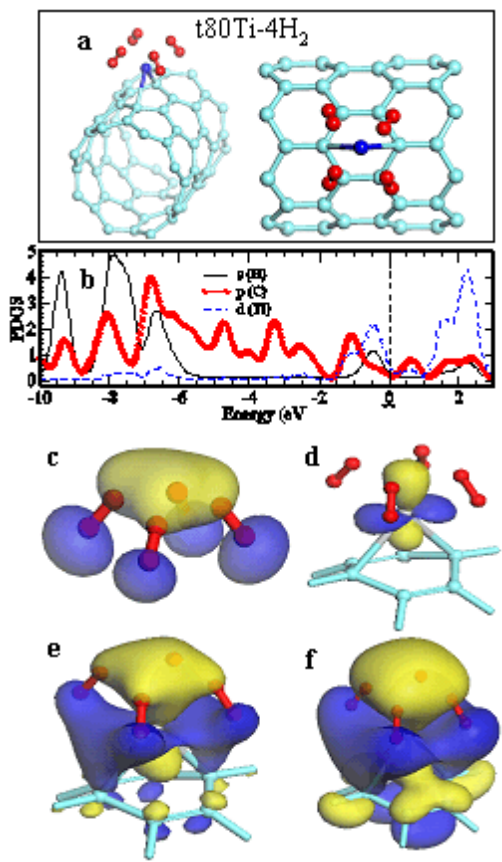}
\caption{
(a) Two different views of the optimized structure of {\bf t80Ti-4H$_{2}$}. 
The relevant structural parameters are: H-H=0.84 \AA, Ti-H=1.9 \AA, 
Ti-C=2.17 \AA, Ti-C'=2.4 \AA.
(b) The projected density of states (PDOS) at 
$\Gamma$-point contributed from Ti, 
four H$_{2}$ molecules and the six carbons of the hexagon on which Ti 
and H$_{2}$ molecules are bonded. 
(c) the $\sigma^{*}$-antibonding orbital of four H$_{2}$ complex,
(d-f) isosurface of the state just below E$_F$ at three different
values; at $\Psi = 0.08$ it is mainly Ti-d orbital, at 
$\Psi=0.04$ the hybridization between d-orbital, two carbon pi-orbital
and 4H$_2$ $\sigma^*$-antibonding is apparent. At
$\Psi=0.02$ it is clear that the other four carbon atoms are also
involved in the bonding.
}
\label{fig2}
\end{figure}

Projecting the plane waves on the pseudo-atomic 
orbitals (Fig.~2b) 
gives detailed information on the nature of the bonding. The density of 
states indicates that Ti $3s-$ and $3p$-orbitals are mostly intact 
and lie below  -20 eV. 
 In the energy range of from -10 eV to zero, 
 Ti contributes only $d$-electrons 
 while Ti 4$s$-electrons are almost absent, 
 probably promoted to Ti $d$-orbitals.
 Fig.~2b shows that the binding state just below $E_{F}$
  has a major contribution 
 from Ti $d$-orbitals, together with the carbon $p$-orbitals 
 and hydrogen $s$-orbitals. In Fig.~2, we showed the isosurface
plot of this state. From this figure it is clear that
a filled Ti $d$-orbital  is 
hybridized simultaneously with 
the  $\sigma*$-antibonding 
molecular orbital of 4H$_{2}$ complex (see Fig.~2c)
and the  $\pi_4^*$-antibonding orbital
of the C6-ring of the SWNT.  The Mulliken analysis
indicate that there is about 1 e transfer to the 4H$_2$ antibonding
orbital. Similarly PDOS in Fig.~2b shows that the
$\sigma$-bonding of the 4$H_2$ group forms a band between
-10 and -6 eV and weakly hybridized with d-orbitals. 
Integrating these peaks over energy and k-points 
suggests that there is about 0.4 e electron transferred to
the Ti empty $d$-orbitals. 
Hence the bonding mechanism for t80Ti-4H$_2$ 
seems to be very similar to Dewar, Chatt, and 
Duncanson model\cite{DCD}, where the interaction has often
viewed as a donation of charge from the highest occupied orbital
of the ligand to the metal empty states and a subsequent back
donation from filled d-orbitals into the lowest unoccupied 
orbital of the ligand. Finally, the second peak around 
-1 eV in PDOS corresponds to the
hybridization of one $d$-orbitals with the $\pi$ orbitals of
the C6-hexagon of SWNT, which is responsible for the
bonding of Ti atom to the nanotube.

In summary, our analysis of PDOS and molecular
orbitals clearly indicates that we need two occupied d-orbitals;
one for molecular bonding of the hydrogens and the other to bind 
the metal to the nanotube. We also expect that the ionization
potential (IP) of the metal is important, which controls the amount
of back charge transfer to the hydrogen antibonding state. 
When a single H$_2$ molecule is introduced to t80Ti,
it seems that Ti is able to donate just enough charge to the
$\sigma^*$-antibonding state, causing dihydrogen to be
unstable against dissociation of H$_{2}$ (Fig. 1a).
However when more hydrogen molecules are added to the system, 
the charge transfer per H$_2$ molecule is not enough to destabilize
the dihydrogen state and therefore the absorption becomes
molecular (Fig. 2a).


From above discussion it is clear that we need both filled
and empty $d$-orbitals for the metal-hydrogen complex formation.
We tried the same thing with alkali and alkaline earth metals,
such as Li, Mg and failed. Heavy transition metals with diffusive
d-orbitals, such as Pt and Pd,  are also not good candidates for 
the molecular absorption. Such metals are known to interact with the
$\sigma*$-antibonding of the hydrogen molecules strongly, destabilizing
dihydrogen structure against classical hydride formation.
Our preliminary results based on a 
structural optimization starting from 
{\bf t80TiH$_2$-3H$_2$} with Ti replaced by Pt/Pd 
indeed indicate that the side hydrogen molecules are not bonded and leave the
 system immediately\cite{tanerprb}. We did observe that two 
 H$_2$ do indeed bind to Pt/Pd forming 
 a PtH$_{4}$ (or PdH$_{4}$) classical hydride cluster, which 
 was not bonded to the nanotube. 
However we expect the light transition metals like Sc and V
to show similar behavior
since they have both occupied and empty d-orbitals with a similar
IP to Ti. We are currently investigating a large number of transition
metals and the results will be published elsewhere\cite{tanerprb}.

It is important to know if the results reported above 
for (8,0) SWNT  hold for other nanotubes and how they depends on
 the chirality and tube radius. Therefore
we have also studied four  (n,n) (n=4,5,6 and 7) and
five  (n,0) (n=7,8,9,10, 11, and 12) nanotubes and 
details will be published elsewhere\cite{tanerprb}. 
Briefly we find that the binding energies of 
TiH$_2$, TiH$_2$-3H$_2$, and Ti-4H$_2$ groups 
have a weak but complicated dependence on
the tube radius, conduction 
and valence band energy levels and band gaps. 
Therefore we did not find a simple trend like 1/R for 
the binding energies. This 
suggests that nanotubes with larger diameter can also show the
similar effect. 
For the largest nanotubes that we have studied,
the binding energy of Ti 
is reduced to 1.8 and 1.6 eV for (12,0) and (7,7), respectively.
The binding energy for the first H$_2$ for (7,7) is 
about 0.66 eV/H$_{2}$, slightly reduced from 
0.83 eV/H$_{2}$ for (8,0) nanotube.  We  find that those nanotubes 
with significant band gaps (like (8,0), (10,0) and (11,0)) 
show the strongest hydrogen absorption. For example (10,0) has 
2.5 eV/4H$_{2}$ binding energy for TiH$_2$-3H$_2$ 
while the binding energy for (11,0) and (8,0) is 
about 2.1 eV/4H$_{2}$.
We also find interesting small differences between (n,n) and 
(n,0) nanotubes.  For example the 
TiH$_2$-3H$_2$ configuration (Fig. 1c) is the ground state 
for the (n,n) nanotubes, while it is  Ti-4H$_2$ (Fig. 2a)
for (n,0) nanotubes. 
In conclusion, the phenomena that {\it a single Ti atom 
absorbed on hexagonal phase of  SWNT can bind up to 
four molecular hydrogen} is a very general and novel result and 
holds for a very large number of nanotubes and possible for 
other carbon-based nanoclusters (i.e. C$_{60}$\cite{c60tihx}).

To this point we have discussed the interaction of H$_{2}$ with a single Ti atom bonded to a nanotube, but clearly one can imagine attaching more Ti to a nanotube, thereby increasing the hydrogen storage capacity. In order to show the feasibility 
 of this approach, we present two simple cases where Ti covers $\frac{1}{4}$ and 
 $\frac{1}{2}$ of the hexagons.  
 The optimized bond-lengths and other parameters of the 
 structures shown in Fig.~3 are very similar to those in the single-Ti case, 
 indicating that the system has the capacity to have many Ti and hydrogen. 
 In fact, these configurations, which have the chemical formulas, 
 C$_{8}$TiH$_{8}$ and C$_{4}$TiH$_{8}$, store 
 approximately 5-wt\% and 8-wt\% hydrogen, respectively.
These numbers are based on the assumption that t80TiH$_2$
group will also release  the hydrogen molecule without difficulty.
This is a good assumption since the binding energy of H$_2$ in
t80TiH$_2$ is about 1/3 of the binding energy of t80Ti.

\begin{figure}
\includegraphics[scale=0.50,angle=0]{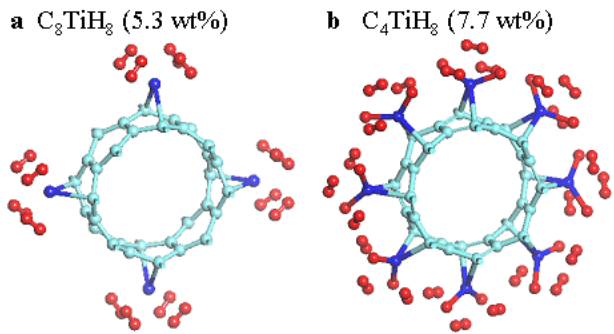}
\caption{
Two high-density hydrogen coverage on Ti-coated (8,0) nanotube. 
}
\label{fig3}
\end{figure}

The calculated binding energy\cite{be} for C$_{8}$TiH$_{8}$  
is 0.43 eV per H$_{2}$, 
which is somewhat reduced from the average binding energy 
of 0.54 eV in {\bf t80Ti-4H$_{2}$}. 
For C$_{4}$TiH$_{8}$, the binding energy is further reduced to 0.18 eV. 
This reduction is due to Ti-Ti interaction which is also responsible
for the increased binding energy of Ti on nanotubes 
(i.e. 2.8 eV/Ti compared to 2.2 eV for a single Ti on a (8,0) SWNT). 
The stability of C$_{4}$TiH$_{8}$ is 
further checked by MD simulations on a $1\times 1\times 2$ supercell, 
which did not show any Ti metal segregation. 
Furthermore, we checked the stability of the system against 
releasing one and eight of the top hydrogen molecules 
(see Fig. 3b), both yielded a binding energy of 0.13 eV/H$_2$.
This is the weakest bond in the system and yet it is 4-5 times stronger
than the van der Walls interactions between hydrogen and SWNT\cite{tanertube}.

In conclusion, using the state-of-the-art
 first-principles total-energy calculations 
we have shown that each Ti atom adsorbed on a SWNT 
can bind up to four hydrogen molecules, a remarkable and totally 
unanticipated finding. 
The mechanism of the bonding is explained by a unique 
hybridization between Ti-$d$ ,
hydrogen $\sigma^*$-antibonding and SWNT LUMO orbitals (Fig.~2).
These results advance our fundamental understanding of dissociative 
and molecular chemisorption of hydrogen in nanostructures, a fundamental
step towards novel materials needed for hydrogen production, storage,
and consumption in the fuel cells. They also  suggest 
a possible method of engineering new nanostructures for high-capacity 
storage and catalyst materials. 

{\bf Acknowledgments:} 
We thank R. Cappelletti, D. A. Neumann and J. J. Rush for 
critical reading of the manuscript and S. Dag, J. \'I\~niguez, 
J. E. Fischer, and T. J. Udovic for discussions. 
This work was partially supported by DOE under Grant 
No.  DEFC36-04-GO14280 and by NSF under Grant No. INT0115021.


\begin{thebibliography}{99}
\bibitem{segregation}
This is due to compatible ionic
radius and the Ti-Ti and Ti-C interactions which are similar in strength.
\bibitem{be} We defined the binding-energy as the energy sum
 of the individual H$_2$ molecules and the tube-Ti 
spin-polarized optimized structure minus the 
energy of the final optimized structure.
\end{thebibliography}
\end{document}